\documentclass[journal,transmag]{IEEEtran}
\usepackage{mathrsfs}
\usepackage{pifont}
\usepackage{bbding}
\usepackage{tipa}
\usepackage{color}
\usepackage{cite}
\usepackage{epsfig}
\usepackage{graphicx}
\usepackage{url}
\usepackage{amsfonts}
\usepackage{multicol}
\usepackage{float}
\usepackage{times}
\usepackage{psfrag}
\usepackage{subfigure}
\usepackage{stfloats}
\usepackage{amsmath}
\usepackage{footnote}
\usepackage{array}
\usepackage{stmaryrd}
\usepackage{amssymb}
\usepackage{epic}
\usepackage{graphicx}
\usepackage{curves}
\usepackage{balance}
\usepackage{algorithm}
\usepackage{algorithmic}
\usepackage{multirow}
\usepackage{amsmath}
\usepackage{xcolor}
\usepackage{graphics}
\usepackage{epstopdf}
\usepackage{enumerate}
\usepackage{geometry}
\floatname{algorithm}{Algorithm}

\newtheorem{assumption}{Assumption}
\newtheorem{lemma}{Lemma}
\newtheorem{definition}{Definition}
\newtheorem{remark}{Remark}
\newtheorem{corollary}{Corollary}

\begin{document}

\title{Dynamic Virtual Resource Allocation for 5G and Beyond Network Slicing }
\author{Fei Song, Jun Li, Chuan Ma, Yijin Zhang, Long Shi, and Dushantha Nalin K. Jayakody Li}

\IEEEtitleabstractindextext{
\begin{abstract}
  The fifth generation and beyond wireless communication will support vastly heterogeneous services and use demands such as massive connection, low latency and high transmission rate. Network slicing has been envisaged as an efficient technology to meet these diverse demands. In this paper, we propose a dynamic virtual resources allocation scheme based on the radio access network (RAN) slicing for uplink communications to ensure the quality-of-service (QoS). To maximum the weighted-sum transmission rate performance under delay constraint, formulate a joint optimization problem of subchannel allocation and power control as an infinite-horizon average-reward constrained Markov decision process (CMDP) problem. Based on the equivalent Bellman equation, the optimal control policy is first derived by the value iteration algorithm. However, the optimal policy suffers from the widely known curse-of-dimensionality problem. To address this problem, the linear value function approximation (approximate dynamic programming) is adopted. Then, the subchannel allocation \emph{Q-factor} is decomposed into the per-slice \emph{Q-factor}. Furthermore, the \emph{Q-factor} and Lagrangian multipliers are updated by the use of an online stochastic learning algorithm. Finally, simulation results reveal that the proposed algorithm can meet the delay requirements and improve the user transmission rate compared with baseline schemes.
\end{abstract}
}

\maketitle

  \section{Introduction}
  Driven by an astounding increase of mobile users and their bandwidth-hungry and varied applications, wireless data traffic will growth exponentially in the next couple years~\cite{3cs},~\cite{1pra}. The fifth generation (5G) and next-generation wireless communication networks will go beyond bringing only high transmission rates for mobile users, as it will support a wider communication ecosystem for the machine-type communications, Internet of Vehicles and Internet of Things \cite{nsm}.
  To promote this evolution, the 5G and beyond wireless communications are envisaged to be the cornerstone for abundant emerging applications~\cite{ai}.
  Due to the diversity of these applications and the complexity of the real environment \cite{6dcu}, the 5G and beyond wireless communications try to guarantee heterogeneous user quality of service (QoS) requirements~\cite{nsb}.
  For instance, it is required to provide extremely high reliability and low latency communications for some vertical industries, like industrial automation control system and Internet of Vehicles, in order to satisfy the stringent QoS demands. Reexamining the networking technologies and network architecture for 5G and beyond networks is necessary according to the varied and customized QoS requirements.
  To establish the service classification, the International Telecommunication Union (ITU) has characterized three types of users~\cite{abs}: 1) Enhanced mobile broadband (eMBB) communications, which demand high transmission rates; 2) Ultra-reliable low-latency communications (URLLC), devised to satisfy stringent latency and reliability requirements; 3) Massive machine type communications (mMTC), aimed to support massive devices, each one uploading very short data packets.

  Network slicing is a vital network architecture innovation in 5G, which is also envisioned to play an important role in the next generation \cite{e2e}, \cite{tpc}. Network slicing permits multiple independent and isolated virtual networks to coexist in the same physical network (PN) infrastructure, the virtual network is defined as slice.
  Software-defined networking (SDN) and network functions virtualization (NFV) are key technologies to implement network slicing for accommodating new services with wide different requirements over the same PN \cite{nas}.
  On the one hand, the slices are established through an abstract set of control logic and resources provided by the SDN controller \cite{raf}. In addition, with SDN the network slicing enables the share of the same PN resources among different tenants.
  On the other hand, NFV is developed to solve the shortage of existing special communication equipment. The core of NFV is virtual network functions, running on virtual machines of general servers without requiring dedicated hardware \cite{sef}.
  Generally, the concept of network slicing prevails due to the following enticing advantages:
  First, network slicing supports multi-tenancy through the virtual networks multiplexing, which leads to the same physical infrastructure can be shared by several virtual network operators \cite{fns}.
  Second, network slicing can realize differentiated service, ensure the service level agreement for every service type \cite{ai}.
  Third, network slicing enhances the adaptability and flexibility of network management as slices can be created as needed and changed on-demand~\cite{nsf}.

  Fig. 1 describes the architecture of a 5G and beyond network scenario with network slicing. In this model, the underlying physical infrastructure are abstracted into network slices, which are managed and coordinated uniformity by orchestrator. The physical resources, such as core networks (CNs), radio access networks (RANs), are divided into several logical parts, thereby forming different network slices based on the user requirements. The slice layer contains multiple slices dedicated to heterogeneous services and runs on the top of underlying layer.
  By providing flexible and scalable network architecture to guarantee various QoS demands of heterogeneous services, RAN slicing is considered to be one of the most promising technologies in 5G and beyond networks.
  RAN slicing can provide customized services for the isolated logical networks by dividing the same PN into multiple isolated logical networks.
  On the basis of sharing PN, RAN slicing is an economic and high-efficient network management scheme.
  A research reported the global capital expenditure and operational expenditure can save nearly 60 billion through RAN sharing by 2021. In practice, the 3GPP has implemented expansive research for 5G network slicing.

   \begin{figure}[H]
   \centering
    \includegraphics[width=3.2in,angle=0]{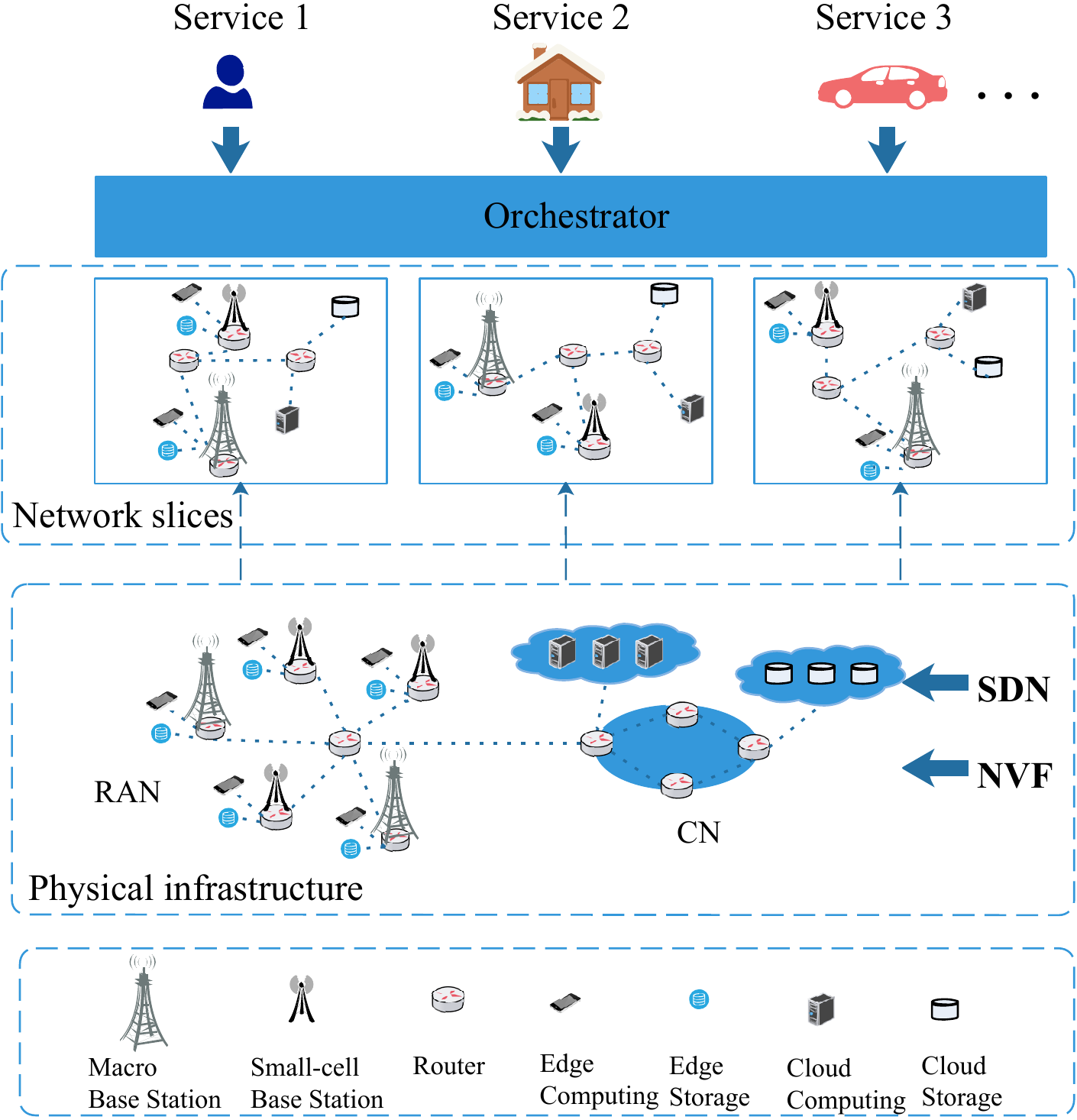}
    \caption{An illustration of network-slicing based 5G and beyond system architecture.}
   \end{figure}

  Due to the properties of RAN slicing, the key problem of RAN slicing is how to flexibly and efficiently allocate the network resources to guarantee the varied QoS demands.
  In RAN slicing, operators allocate resources efficiently and flexibly at the network level according to different performance requirements of users. These resources include not only wireless resources, but also computing and cache resources.
  Resource allocation technology can realize the efficient utilization and improve the utilization rate of resources, and integrate different slices.
  Depending on the environment the resource allocation of network slicing is classified as static allocation and dynamic allocation \cite{joo}.
  Static resource allocation means that, after the resource allocation and mapping strategy of the network slicing is determined, it will remain the same regardless of how the environment changes.
  Dynamic resource management \cite{irs} refers to the ability to sense the changes in the environment, and then dynamically adjust resource allocation strategies to optimize the quality of communication services.

  Network slicing has also attracted widespread research interest in academia.
  The authors of~\cite{uac} studied the resource management of wireless access network slices from the perspective of user access control and wireless bandwidth allocation. The optimization problem of this paper is to minimize bandwidth consumption under the QoS and resource constraints. Finally, the Lagrangian decomposition theory is used to solve the optimization problem.
  In \cite{ocp}, a cell planning scheme for network slicing was proposed to maximize the resource efficiency under different QoS requirements of different mobile services for wireless communications.
  This scheme jointly optimized the inter-slice resource allocation, the distribution of cells operated on different slices, and the users allocation for cells.
  Caballero \emph{et. al.}~\cite{nsfg} proposed a network slicing framework, including admission control, resource management and user dropping.
  However, the wireless channel set in \cite{uac} and \cite{ocp} are fixed. In addition, the resource allocation scheme in \cite{nsfg} was fixed and will not change with the topology. In fact, the wireless channel is always interfered by the environment and the channel quality changes dynamically with time. And the changes in the environment may cause the server to stop working when the policy is fixed.
  Although the above papers studied and analyzed the resource allocation and optimization problem of network slicing, with the exponential growth of mobile terminal data traffic, the increase of device connections and the rise of multimedia and other emerging services, the static resource allocation scheme can not change the resource allocation strategy in real time. Not only the resource utilization is low, but also the performance requirements of all slices cannot be met, so the static resource allocation method cannot meet the current network environment.

  In fact, the wireless channel is always interfered by the environment and the channel quality changes dynamically with time. For this reason, the static resource allocation cannot meet the dynamic performance requirements.
  Driven by this issue, \cite{do1} investigated the optimal dynamic resource allocation and mode selection to minimize the average delay with a dropping probability constraint in D2D communications. This paper formulated the resource control problem in D2D communications into a constrained Markov decision process (CMDP).
  The authors of \cite{dpc} focused on the optimization of dynamic power control policy. The optimization problem was formulated as a CMDP to minimize the weighted sum of average delay and average transmit power.
  However, aforementioned works in \cite{do1} and \cite{dpc} only investigated one type of service requirement for resource allocation.

  In general, the resources need to be allocate dynamically according to the different QoS, and network slicing can meet this demand. A systematic approach of dynamic resource allocation is via MDP and dynamic programming techniques.
  \cite{dns} focused on the fog computing system, wherein a orchestrator coordinated workload distribution among local fog nodes. The energy and computing resources were dynamically allocated to slices.
  In particular, the partially observable MDP (POMDP) was developed in this paper to maximum the total amount of offloaded workload.
  The authors of~\cite{avr} proposed a virtual resource scheduling scheme in 5G slicing network with the non-orthogonal multiple access system. To solve the problem of virtual resource allocation of downlink RAN slicing, the problem was described as a CMDP aiming at maximizing the total user rate.

  Therefore, we focus on the dynamic virtual resource allocation for RAN slicing aided uplink communication systems.
  The subchannel allocation and power control policy is proposed to maximum transmission rate with the heterogeneous dropping probability and delay constrains.
  The rate maximization problem is formulated as an infinite-horizon average reward CMDP.
  The conventional value iteration and policy iteration will leads to the curse of dimensionality.
  To reduce the complexity, we use linear value function approximation and distributed online stochastic learning to simplify the optimization problem with the subchannel allocation and power control.
  The contributions of this paper are summarized as follows.
  \begin{itemize}
    \item A dynamic virtual resource allocation algorithm is proposed for CMDP-based RAN slicing. The resource allocation scheme is dynamically adjusted to power control and subchannel allocation according to the continuous interaction process with external environment. Due to the specific structure of this problem, we derive a reduced-state equivalent Bellman equation.
    \item To further reduce the computational complexity, we use approximate dynamic programming to develop the proposed algorithm. Then, we propose an equivalent Bellman equation in terms of subchannel allocation \emph{Q-factor} to solve the CMDP. Especially, the \emph{Q-factor} is approximated as the sum of per-slice \emph{Q-factor}.
        Furthermore, we put forth a distributed online learning algorithm based on channel state information (CSI), queue state information (QSI) and energy state information (ESI) to optimize the per-slice \emph{Q-factor} and the Lagrange multipliers (LM).
  \end{itemize}

  \begin{table*}[t]
  \newcommand{\tabincell}[2]{\begin{tabular}{@{}#1@{}}#2\end{tabular}}
  \centering
  \caption{Main notations and their definitions}
  \begin{tabular}{|c|c|c|}
    \hline
    \multicolumn {1}{|c|}{Category} & \multicolumn {1}{|c|}{Notations} & \multicolumn {1}{c|}{Definitions}                            \\
    \hline
    \multirow{1}*{Constant}
    &$B$  &Channel bandwidth     \\
    \cline{2-3}
    &$D_m^{max}$  &The maximum delay of slice $m$    \\
    \cline{2-3}
    &$B^{\rm{Q}}_m, \beta_m, \lambda^{\rm{Q}}_m$  &\tabincell{c}{The buffer capacity of UE in slice $m$, the probability that the UE in slice $m$ \\ generate a new queue task, the average data arrival rate of the UE in slice $m$ }    \\
    \cline{2-3}
    &$B^{\rm{E}}_m, \lambda^{\rm{E}}_m$  &\tabincell{c}{The battery capacity of UE in slice $m$, \\ the average energy arrival rate of the UE in slice $m$ }   \\
    \hline
    \multirow{1}*{Variable}
    &$M$, &The number of network slices                            \\
    \cline{2-3}
    &$N$    & The number of subchannels        \\
    \cline{2-3}
    &$k_m$  & The active UEs of slice $m$                                          \\
    \cline{2-3}
    &$g^n_{i,m}, h^n_{i,m}$           &\tabincell{c}{The fading and the channel from the UE $i$ in slice $m$ \\ connect with subchannel $n$ to the BS}   \\
    \cline{2-3}
    &$\phi, P_{i,m}$       &\tabincell{c}{The FPC factor, the transmit power of the UE $i$ in slice $m$  }   \\
    \cline{2-3}
    &$R_{i,m}$, $R_{m}$  &\tabincell{c}{The transmission rate of the UE $i$ in slice $m$, the total rate of the slice $m$} \\
    \cline{2-3}
    &$D_{i,m}$, $\bar{D}_{m}$  &\tabincell{c}{The delay of the UE $i$ in slice $m$, the average delay of the slice $m$} \\
    \cline{2-3}
    &$Q_{i,m}, A^{\rm{Q}}_{i,m}, L^{\rm{Q}}_{i,m}$   &\tabincell{c}{The queue length, the amount of data arrived \\ and the amount of transmitted packets of UE $i$ in slice $m$ }\\
    \cline{2-3}
    &$E_{i,m}, A^{\rm{E}}_{i,m}, L^{\rm{E}}_{i,m}$   &\tabincell{c}{The energy length, the amount of energy arrived  \\ and the amount of transmission energy consumption of UE $i$ in slice $m$}  \\
    \hline
    \multirow{1}*{State}
    &$\mathbf{S}_t$  &The global system state at time slot $t$, $\mathbf{S}_t=(\mathbf{H}_t, \mathbf{Q}_t, \mathbf{E}_t)$ \\
    \cline{2-3}
    &$\mathbf{H}_t$  &The channel at time slot $t$ \\
    \cline{2-3}
    &$\mathbf{Q}_t$  &The queue at time slot $t$ \\
    \cline{2-3}
    &$\mathbf{E}_t$  &The energy at time slot $t$ \\
    \hline
    \multirow{1}*{Action}
    &$\Omega = (\Omega_c, \Omega_p)$  & A stationary subchannel allocation and power control policy  \\
    \cline{2-3}
    &$\Omega_c(s) = \{c_{i,m}^n\}_{i\in k_m, m\in M, n\in N}$,  &The adjustment action of subchannel allocation \\
    \cline{2-3}
    &$\Omega_p(s) = \{\phi_{i,m}\}_{i\in k_m, m\in M}$  &The adjustment action of power control \\
    \hline
  \end{tabular}
  \end{table*}

  The rest of the paper is organized as follows. The system model is briefly described in Section II. The problem formulation and optimal solution are proposed in Section III, t. In Section IV, we develop a low-complexity distributed online learning algorithm. Simulation results are shown in Section V. Finally, Section VI concludes this paper. The main notations used in this paper are listed in Table I.

  \section{System Model}
  In this section, we first present physical layer model, queues dynamic model and energy dynamic model of RAN slicing, and then elaborate the subchannel allocation and power control policy and data dropping and delay constraint.

  \subsection{Physical Layer Model}

  Here, we develop a uplink wireless communication scenario for an RAN slicing consisting of a single BS and heterogeneous UEs.
  The network first determines the service types of user equipments (UEs), and then UEs are assigned to the related slices based on the QoS demands.
  The time dimension is divided into multiple time slots and each slot lasts $\Delta$ seconds. Given the communication scenario for RAN-slicing uplink wireless communication, in which three types of UEs (eMBB, mMTC and URLLC) access the same BS, as shown in Fig. 2.

  \begin{figure}[H]
   \centering
    \includegraphics[width=2.8in,angle=0]{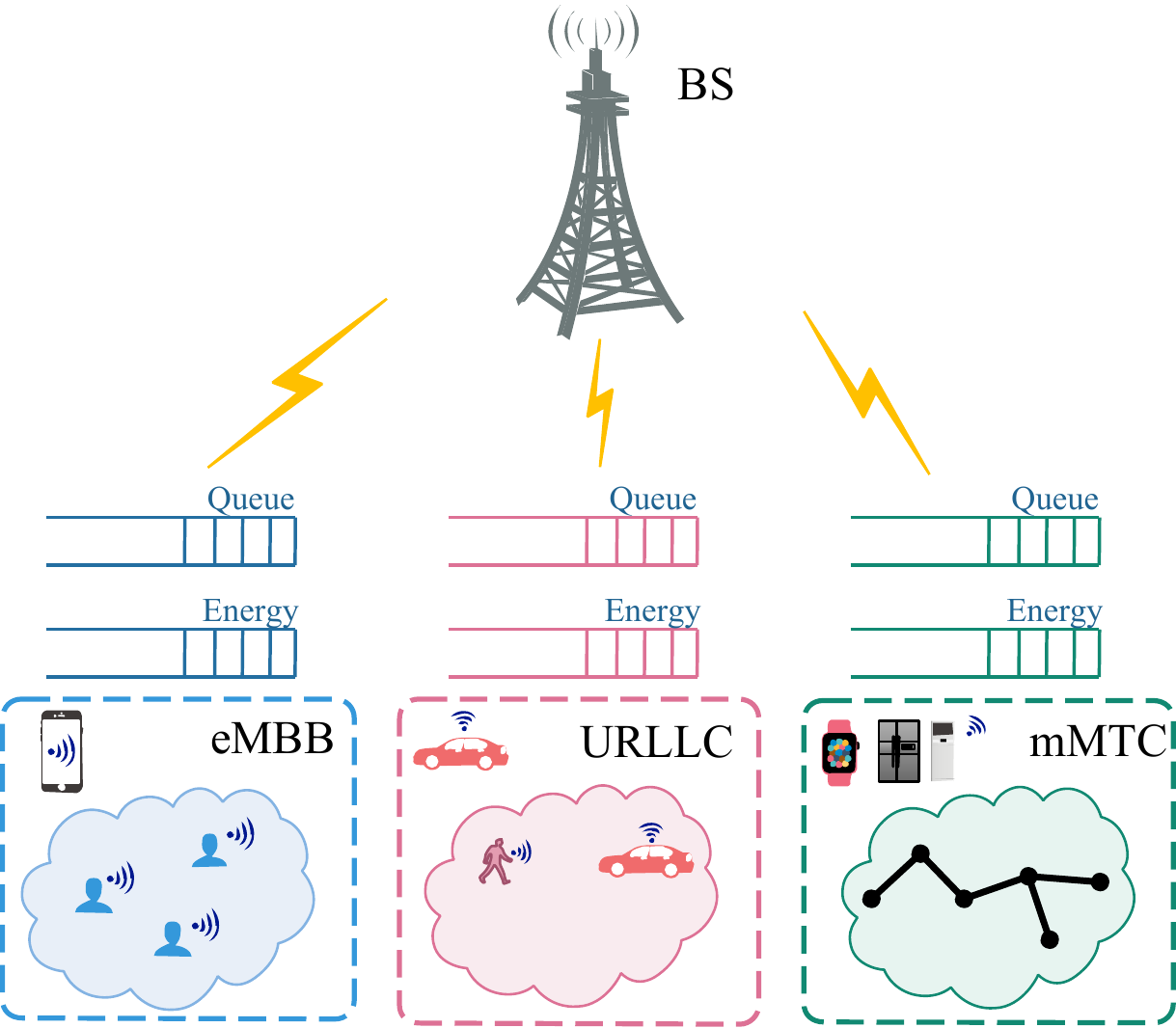}
    \caption{ System scenario for RAN-slicing uplink wireless communication, with three example network slices.}
   \end{figure}

  The wireless resources are divided into $M$ slices to support $M$ different types of services,
  each slice accommodates $k_m$ active UEs, $m\in\{1,2,\cdots,M\}$.
  The BS locate in the center and the UEs are uniformly distributed in a disk with radius $D$. The whole uplink bandwidth $B$ of the BS is divided into $N$ subchannels.

  We focus on the instantaneous channel gain \cite{5pc}, which consists of the fast-fading and path loss on each subchannel $n$.
  Let $g$ be the fast-fading, and $PL$ is the path loss.
  The channel state between the UE in slice $m$ and the subchannel $n$ of BS is denoted by $h^n_{i,m} \in \mathcal{H}$, where $i$ is the UE index in slice $m$ and $\mathcal{H}$ denotes the finite discrete complex channel state space. Let $h^n_{i,m}=g^n_{i,m}\sqrt{PL_{i,m}}=g^n_{i,m}\sqrt{Ad_{i,m}^{-\alpha}}$, where $g^n_{i,m}$ is the fading that the UE $i$ in slice $m$ connect with the subchannel $n$, $d_{i,m}$ denotes the distance from the UE $i$ in the slice $m$ to the BS, $\alpha$ is the path loss factor, and $A$ represents the free space power received at the reference distance $d_{i,m}=1\rm{m}$. We have the following assumptions on CSI.

  \begin{assumption}
  The fading follows a Rayleigh distribution with variance $\sigma_g^2$ and mean $\bar{g}$, and stays the same in a time slot. The fading distribution of each UE is independent identically distributed (i.i.d.) for each time slot and independent with respect to each subchannel \cite{2dc}. During the communication session, the path loss stays the same. Therefore, the CSI stays the same in a time slot. \hfill $\blacksquare$
  \end{assumption}

  We define the transmit power for UE $i$ in slice $m$ as $P_{i,m}$. Given the baseline power $P_m$ of slice $m$, $P_{i,m}$ is given by
  \begin{align}
  \label{equ:1}
  P_{i,m}=P_m(Ad_{i,m}^{-\alpha})^{-\phi},
  \end{align}
  where $\phi$ is the fractional power control (FPC) factor.

  Let $c^n_{i,m}\in\{0,1\}$ denote the subchannel allocation for the UE $i$ in slice $m$ on the $n$th subchannel. Concretely, $c^n_{i,m}=1$ means that the $n$th subchannel is used by the UE $i$ in slice $m$, and $c^n_{i,m}=0$ otherwise. Specifically, considering that the UE $i$ in slice $m$ is scheduled on subchannel $n$, the received signal at the BS from the UE $i$ in slice $m$ on subchannel $n$ is given by
  \begin{align}
   y_{i,m}^{n}=h^n_{i,m} {\sqrt {P_{i,m}} {x_{i,m}}}+\omega_n.
  \end{align}
  Here, $x_{i,m}$ is the transmit signal of UE $i$ in slice $m$. $\omega_n$ denotes a zero-mean complex additive white Gaussian noise (AWGN) random variable on the subchannel $n$ with variance ${\sigma}_n^2$.

  Hence, the transmission rate of the UE $i$ in slice $m$ on subchannel $n$ is given by
  \begin{align}
   {R_{i,m}^n} = B{\log _2}\left( {1 + {\frac{P_{i,m}|{h_{i,m}^n}{|^2}} {\sigma_n^2 }}} \right).
  \end{align}

  \subsection{Queues Dynamic Model}
  Each UE continuously buffer the arriving packets by maintains a traffic queue.
  To reduce the queuing time waiting in the buffer before transmission, it is crucial to properly schedule the queues to maximize the system capacity.
  The generation of new queue task for each UE in slice $m$ follows independent Bernoulli distribution with probability $\beta_m$, which represents the probability of UEs in slice $m$ that generate new queue tasks at the current time slot.
  In addition, the packet arrival process of UE in slice $m$ is assumed to i.i.d. for each time slot and follows Poisson distribution with the average arrival rate $\lambda_m^{\rm{Q}}$.
  To be more formal, we assume that the packet arrivals take place around the slot boundaries.

  In time slot $t$, the amount of data arrived at UE is denoted as $A^{\rm{Q}}_{i,m}(t)$ and the current queue length of UE $i$ in slice $m$ is denoted as $Q_{i,m}(t)$.
  The subsequent arriving packet will be dropped if the queue length reached the buffer capacity $B^{\rm{Q}}_m$. Hence, the queue dynamics can be expressed as
  \begin{align}
  \label{queue}
  &Q_{i,m}(t+1)\nonumber \\
  &=\min\{[Q_{i,m}(t)+A^{\rm{Q}}_{i,m}(t)-L^{\rm{Q}}_{i,m}(t)]^+, B^{\rm{Q}}_m\},
  \end{align}
  where $x^+ \triangleq \max\{x,0\}$, and $L^{\rm{Q}}_{i,m}(t)= R_{i,m}(t)\Delta$ denotes the amount of transmitted packets of UE $i$ in slice $m$ during time slot $t$. Note that $A^{\rm{Q}}_{i,m}(t)=0$ when the new queue task is not generated in time slot $t$.

  Define the average queue length of UE $i$ in slice $m$ is $\overline{Q_{i,m}}$, which is given by
  \begin{align}
   \overline{Q_{i,m}}=\mathop {\lim }\limits_{T \to \infty } \frac{1}{T}\sum\limits_{t = 0}^T E^\Omega[Q_{i,m}(t)] = E^\Omega[Q_{i,m}].
  \end{align}
  where $E[\cdot]$ means the expectation operation.

  \subsection{Energy Dynamic Model}

  Define the energy arriving at UE in time slot $t$ as $A^{\rm{E}}_{i,m}(t)$. $B_m^{\rm{E}}$ denotes the battery capacity. The energy arrival $A^{\rm{E}}_{i,m}(t)$ is assumed to i.i.d. for each scheduling slot and independent follows a general distribution with average arrival rate $\lambda^{\rm{E}}_m$. At time slot $t$, $L^{\rm{E}}_{i,m}(t)$ unit of energy is consumed for packet transfer. Therefore, the energy dynamics can be expressed as
  \begin{align}
  \label{energy}
  &E_{i,m}(t+1) \nonumber \\
  &=\min\{[E_{i,m}(t)+A^{\rm{E}}_{i,m}(t)-L^{\rm{E}}_{i,m}(t)]^+, B_m^{\rm{E}}\},
  \end{align}
  where $L^{\rm{E}}_{i,m}(t)= P_{i,m}(t)\Delta$ denotes transmission energy consumption of UE $i$ in slice $m$ at time slot $t$.

  \subsection{Subchannel Allocation and Power Control Policy}
  Due to the system states change dynamically with time, the resource manager dynamically adjust subchannel allocation and power control strategy in real-time based on global CSI (GCSI), global QSI (GQSI) and global ESI (GESI). At the beginning of the $t$th time slot, the resource manager determines the subchannel allocation, and perform power control in terms of a stationary subchannel allocation and power control policy described as follow.

  \begin{definition}
   The mapping from the system state $\mathbf{S} \in \mathcal{S}$ to the subchannel allocation and power control actions is regarded as a stationary subchannel allocation and power control policy $\Omega = (\Omega_c, \Omega_p)$, where $\Omega_c(\mathbf{S}) = \{c_{i,m}^n\}_{i\in k_m, m\in M, n\in N}\in \mathcal{C}$, and $\Omega_p(\mathbf{S}) = \{\phi_{i,m}\}_{i\in k_m, m\in M} \in \mathcal{P}$.
  Based on \eqref{equ:1}, the FPC factor $\phi$ can adjust the transmit power of UEs to meet the slice QoS requirements. Therefore, the action of power control is given by $\phi_{i,m}\in (0,1]$.  \hfill $\blacksquare$
  \end{definition}

  Note that the instantaneous transmission rate is influenced by the subchannel allocation and power control at time slot $t$, the total rate of UE $i$ in slice $m$ is given by ${R_{i,m}}(t) = \sum\limits_{n=1}^N c_{i,m}^n(t){R_{i,m}^n}(t)$.
  Hence, the total transmission rate for slice $m$ is given by
  \begin{align}
   {R_m}\!(t)\! &=\! \sum\limits_{n = 1}^{N} \sum\limits_{i = 1}^{{k_{m}}} c_{i,m}^n(t) {R_{i,m}^n}(t)  \nonumber \\
   &=\! \sum\limits_{n = 1}^{N} \sum\limits_{i = 1}^{{k_{m}}}\! {c^n_{i,m}(t)}B{\log _2}\left(\! {1 \!+ {\frac{{P_{i,m}(t)|{h_{i,m}^n}(t){|^2}}} {{\sigma_n^2 }(t)}}} \!\right).
  \end{align}

  In order to ensure that transmission is not interrupted by energy exhaustion, the power control policy $\Omega_p$ satisfies the power consumption constraint
  \begin{align}
  E_{i,m}(t)> \min\{P_{i,m}(t){\frac{Q_{i,m}(t)}{R_{i,m}(t)}}, P_{i,m}(t)\cdot \Delta\}.
  \end{align}

  Therefore, the UE upload queue tasks only when energy constraints are met. In this study, we redefine the immediate transmission rate as
  \begin{align}
  {R_{i,m}(t)} = \left\{
  {\begin{array}{*{20}{c}}
  {{R_{i,m}(t)},}&\text{if (8) is satisfied, }\\
  {0,}&{\text{otherwise.}}
  \end{array}} \right.
  \end{align}

  \subsection{Data Dropping and Delay}
  We consider the dropping probability in the system model when the traffic queue buffer size is limited \cite{do1}. At this time, the packet discard is inevitable when the buffer overflows. For the traffic queue with limited buffer size $B_Q$, the dropping probability for UE $i$ in slice $m$ can be written as
  \begin{align}
  \overline{dr_{i,m}}&= 1 - {\frac{{{\text{Average \# of data transmitted in a time slot}}}}  {{\text{Average \# of data arrived in a time slot}}}} \nonumber \\
 &= 1 - {\frac{{\overline{ R_{i,m}} \cdot \Delta }}  {\lambda_m^Q}}.
  \end{align}

  Considering the impact of packet dropping, the Little's Law is extended to a situation with packet dropping \cite{asd}. The average delay of UE $i$ in slice $m$ is given by
  \begin{align}
  \overline{D_{i,m}}= \frac{{\overline{Q_{i,m}}}} {{(1-\overline{dr_{i,m}})\lambda_m^Q}} = \frac{{\overline{Q_{i,m}}}} {\overline{ R_{i,m}} \cdot \Delta}.
  \end{align}

   The average delay of the slice $m$ can be written as
  \begin{equation}
  \overline {D_{m}}= \sum\limits_{i = 1}^{k_m} {\overline{D_{i,m}}} .
  \end{equation}

  \section{Problem formulation and Optimal Solution}

  \subsection{Dynamic Transition of System State}
  The current system state depends on the state and action of the previous time slot at each time slot. Define the network state as $\mathbf{S}(t)$, and it changes over time slots $t=\{1,2,\cdots\}$. $\mathcal{S}$ denote the state space. Let $\mathbf{S}(t)=(\mathbf{H}(t), \mathbf{Q}(t), \mathbf{E}(t))$ at time slot $t$, where the $\mathbf{H}(t)$ describes the channel gain state, $\mathbf{Q}(t)$ describes the queue state, and $\mathbf{E}(t)$ describes the energy state. All the elements in each set are finite discrete values.

  Furthermore, the $\mathbf{H}(t)$ is defined as $\mathbf{H}(t)=\{\mathbf{H}^n_{m}(t) \}_{m\in M, n\in N}$, where $\mathbf{H}^n_{m}(t) = \{h^n_{i,m}(t)\}_{ i \in k_{m}}$,
  $\mathbf{Q}(t)=\{\mathbf{Q}_{m}(t) \}_{m\in M}$ with $\mathbf{Q}_{m}(t)=\{Q_{i,m}(t)\}_{ i \in k_{m}}$,
  and $\mathbf{E}(t)=\{\mathbf{E}_{m}(t) \}_{m\in M}$ with $\mathbf{E}_{m}(t) = \{E_{i,m}(t)\}_{ i \in k_{m}}$.

  \vspace{1mm}
  We represent the channel model, queue model and energy model as discrete-time Markov chain (DTMC). Therefore, $\{\mathbf{S}(t)\}$ for given $\Omega$ is Markovian, and the state transition probability is given by
  \begin{small}
 \begin{align}
 \label{equ:6}
 \Pr&\{\mathbf{S}(t+1)|\mathbf{S}(t),\Omega(\mathbf{S}(t))\} \nonumber \\
 &=\Pr\{\mathbf{H}(t+1)|\mathbf{S}(t),\Omega(\mathbf{S}(t))\}\Pr\{\mathbf{Q}(t+1)|\mathbf{S}(t),\Omega(\mathbf{S}(t))\} \nonumber \\
 &\;\;\;\;\times\Pr\{\mathbf{E}(t+1)|\mathbf{S}(t),\Omega(\mathbf{S}(t))\} \nonumber \\
 &=\Pr\{\mathbf{H}(t+1)\}\Pr\{\mathbf{Q}(t+1)|\mathbf{S}(t),\Omega(\mathbf{S}(t))\} \nonumber \\
 &\;\;\;\;\times\Pr\{\mathbf{E}(t+1)|\mathbf{S}(t),\Omega(\mathbf{S}(t))\}.
 \end{align}
 \end{small}

  \vspace{-5mm}
  Firstly, the queue state transition probability $\Pr\{\mathbf{Q}(t+1)|\mathbf{S}(t),\Omega(\mathbf{S}(t))\}$ is derived. Given the state $\mathbf{S}(t)$ and policy $\Omega(\mathbf{S}(t))$, the conditional probability of $Q_{i,m}(t+1)$ based on \eqref{queue} is given by
  \begin{align}
  \Pr&\{Q_{i,m}(t+1)|\mathbf{S}(t),\Omega(\mathbf{S}(t))\} \nonumber \\
  &=\!\beta_m\Pr(Q_{i,m}(t)\!=\!q)\!=\!\beta_m\frac{{(\lambda _m^Q)^q}}{{q!}}\exp \left( { - {\lambda^Q _m}} \right).
  \end{align}

  The $\Pr\{\mathbf{Q}(t+1)|\mathbf{S}(t), \Omega(\mathbf{S}(t))\}$ can be written as the product of $\Pr\{Q_{i,m}(t+1)|\mathbf{S}(t), \Omega(\mathbf{S}(t))\}$ over all UEs as
  \begin{align}
  \Pr&\{\mathbf{Q}(t+1)|\mathbf{S}(t), \Omega(\mathbf{S}(t))\} \nonumber\\
  &= \prod\limits_{m = 1}^{M} \prod\limits_{i = 1}^{k_m} \Pr\{Q_{i,m}(t+1)|\mathbf{S}(t), \Omega(\mathbf{S}(t))\}.
  \end{align}

  Similarly, the energy state transition probability $\Pr\{\mathbf{E}(t+1)|\mathbf{S}(t),\Omega(\mathbf{S}(t))\}$ is derived. Given the state $\mathbf{S}(t)$ and policy $\Omega(\mathbf{S}(t))$, the conditional probability of $E_{i,m}(t)$ based on \eqref{energy} is given by
  \begin{align}
  \Pr&\{E_{i,m}(t+1)|\mathbf{S}(t),\Omega(\mathbf{S}(t))\} \nonumber \\
  &=\Pr(E_{i,m}(t)=e)=\frac{{(\lambda^E_m)^e}}{{e!}}\exp \left( { - {\lambda^E _m}} \right).
  \end{align}

  The energy state transition probability $\Pr\{\mathbf{E}(t+1)|\mathbf{S}(t), \Omega(\mathbf{S}(t))\}$ can be written as the product of $\Pr\{E_{i,m}(t+1)|\mathbf{S}(t), \Omega(\mathbf{S}(t))\}$ over all UEs as
  \begin{align}
  \Pr&\{\mathbf{E}(t+1)|\mathbf{S}(t), \Omega(\mathbf{S}(t))\} \nonumber \\
  &= \prod\limits_{m = 1}^{M} \prod\limits_{i = 1}^{k_m} \Pr\{E_{i,m}(t+1)|\mathbf{S}(t), \Omega(\mathbf{S}(t))\}.
  \end{align}

  \subsection{Problem Formation}
  The objective is maximizing the average weighted-sum transmission rate by optimizing the subchannel allocation and power control policy under the heterogeneous delay constraints.
  It is common knowledge that desirable data transmission rate is a key factor to improve user satisfaction \cite{4osc}.
  The induced Markov chain $\{\mathbf{S}(t)\}$ is ergodic and has a unique steady state distribution $\pi_s$ given a unichain policy $\Omega$.
  Consequently, the average reward for $m$th slice can be derived as
  \begin{align}
  \overline{R_m}(\Omega)&=\mathop {\lim \sup} \limits_{T \to \infty } \frac{1}{T} \sum\limits_{t = 0}^T {\mathbb{E}}^{\Omega} [R_{m}(t)] \nonumber \\
  &= {\mathbb{E}}^{\pi({\Omega})} [R_{m}].
  \end{align}

  \emph{Problem 1:} The objective of the dynamic virtual resource allocation is to maximize the reward function subject to the constraints, which can be expressed as
  \begin{align}
  & \max  \sum\limits_{m = 1}^M \omega_m {R_m}(\Omega) = \mathop {\lim} \limits_{T \to \infty } \frac{1}{T} \sum\limits_{t = 0}^T {\mathbb{E}}^{\Omega} [d(\mathbf{S}(t), \Omega(\mathbf{S}(t)))], \\
  &s.t.\;  {\overline{D_{m}}}\leq D_m^{\max}\;\;\;\;\;   \forall m={1,2\ldots, M},
  \end{align}
  where $\omega_m$ is the weight for slice $m$, $D_m^{\max}$ is the maximum delay that slice $m$ can tolerate, and $d(\mathbf{S}(t), \Omega(\mathbf{S}(t))) = \sum\limits_{m = 1}^M \omega_m R_{m}(t)$.

  \subsection{Constrained MDP}
  The proposed \emph{Problem 1} is considered as an infinite-horizon average reward CMDP.
  All elements in each set are finite discrete values.
  Based on the global system state $\mathbf{S} = (\mathbf{H}, \mathbf{Q}, \mathbf{E}) \in \mathcal{S}$, where $\mathcal{S} = \mathcal{H}\times \mathcal{Q}\times \mathcal{E}$,  the channel allocation and power control policy is optimized. Hence, the \emph{Problem 1} is a CMDP with the following definition:
  \begin{itemize}
    \item State Space: $\{(\mathbf{H},\mathbf{Q},\mathbf{E}): \forall \mathbf{H}\in \mathcal{H},\mathbf{Q}\in \mathcal{Q}, \mathbf{E}\in \mathcal{E}\}$
    \item Action Space: $\{\Omega(\mathbf{H},\mathbf{Q},\mathbf{E}): \forall \mathbf{H}\in \mathcal{H}, \mathbf{Q}\in \mathcal{Q}, \mathbf{E}\in \mathcal{E}\}$, where $\Omega = (\Omega_c, \Omega_p)$ is described in \emph{Definition 1}.
    \item State Transition Probability: $\Pr[\mathbf{S}'|\mathbf{S},\Omega(\mathbf{S})]$ can be formulated as (\ref{equ:6}).
    \item Per-stage Reward Function: $d(\mathbf{S}(t), \Omega(\mathbf{S}(t))) = \sum\limits_{m = 1}^M \omega_m R_{m}(t)$.
  \end{itemize}

  The MDP problem is convex programming problem, which is shown in \cite{do1}. Therefore, the constraints of \emph{Problem 1} can be converted into an expanded objective function with a weighted-sum of the constraint functions by the Lagrangian duality method.
  The dual problem of the CMDP in \emph{problem 1} is considered. For any given nonnegative Lagrangian multipliers (LMs) $\mathbf{\eta}=\{{\eta_m}|m\in M\}$, the Lagrangian function of \emph{Problem 1} is given by
  \begin{align}
  \label{cmdp}
  L(\Omega,\mathbf{\eta}) = \mathop {\lim} \limits_{T \to \infty } \frac{1}{T} \sum\limits_{t = 0}^T {\mathbb{E}}^{\Omega} [g(\mathbf{\eta}, \mathbf{S}(t), \Omega(\mathbf{S}(t)))],
  \end{align}
  where
  \begin{align}
  \label{gx}
  g(\mathbf{\eta}, \mathbf{S}, \Omega(\mathbf{S}))\!= \! \sum\limits_{m=1 }^M \left(\omega_m R_m \!+ \eta_m (\overline{D_m}\! \!-D_m^{\max})\right).
  \end{align}

  As a result, we divide the \emph{problem 1} into two subproblems, which is given by
  \begin{align}
  &{\text{Subproblem 1-1}}: G(\eta)= \max\limits_{\Omega} L(\Omega, \eta) \nonumber, \\
  &{\text{Subproblem 1-2}}: G(\eta^*)= \min\limits_{\eta} G(\eta) \nonumber,
  \end{align}
  where Subproblem 1-1 and Subproblem 1-2 are the Lagrange dual function and the dual problem, respectively.

  Consequently, we can get the optimizing policy by recursively solving the Bellman equation~\cite{dp}, which is given by
  \begin{align}
  \label{equ:v}
  &\theta+ V(\mathbf{S}) \nonumber\\
  &=\! \max\limits_{\Omega(\mathbf{S})} \{g(\eta, \mathbf{S}, \Omega(\mathbf{S}))\! +\! \sum\limits_{\mathbf{S}'}\!\Pr[\mathbf{S}'|\mathbf{S}, \Omega(\mathbf{S})]V(\mathbf{S}') \},
  \end{align}
  where $\theta$ is the per-period optimal average reward for the steady-state system; $V(\mathbf{S})$ represents the value function, which represents the average reward attained according to the control policy $\Omega$ of each state $\mathbf{S}$;
  $\Omega(\mathbf{S})= (\Omega_c(\mathbf{S}), \Omega_p(\mathbf{S}))$ are the subchannel allocation and power control policy performed in state $\mathbf{S}$; $g(\eta, \mathbf{S}, \Omega(\mathbf{S}))$ in \eqref{gx} is the per-stage reward function under the $\mathbf{S}$ and $\Omega(\mathbf{S})$.
  If $(\theta, \{V(\mathbf{S})\})$ can satisfy the fixed-point equations in (\ref{equ:v}), the $\theta = \max_\Omega L(\Omega, \eta)$ will be the optimal solution
  in \emph{Problem 1}.
  In addition, the stationary policy is optimal when $\Omega^\ast(\eta) = (\Omega^\ast_c, \Omega^\ast_p)$ obtains the maximum of the R.H.S. of \eqref{equ:v} for all states.

  \subsection{Equivalent Bellman Equation}

  Using the i.i.d. property of the CSI, the state space can be reduced.
  According to the partial system state $(\mathbf{Q}, \mathbf{E})$, a reduced-state Bellman equation from (23) is derived, instead of studying the complete system state $\mathbf{S}$.
  Specifically, the conditional actions of a policy $\Omega$ is given by following.

  \begin{definition}
  Given a control policy $\Omega = (\Omega_c, \Omega_p)$, $\Omega(\mathbf{Q}, \mathbf{E}) = \{(\mathbf{c}, \mathbf{p}) = (\Omega(\mathbf{S})): \mathbf{S}=(\mathbf{H}, \mathbf{Q}, \mathbf{E}) \forall \mathcal{H}\}$ is defined as the collection of actions $(\mathbf{c}, \mathbf{p})$ for all possible CSI $\mathbf{H}$ conditioned on a given QSI and ESI pair $(\mathbf{Q}, \mathbf{E})$.
  Thus, the policy $\Omega$ is the union of all conditional actions, i.e., $\Omega = \bigcup_{(\mathbf{Q}, \mathbf{E})}\Omega(\mathbf{Q}, \mathbf{E})$ . \hfill $\blacksquare$
  \end{definition}

  By solving a reduced-state equivalent Bellman equation, we can attain the optimal control policy in \emph{Problem 1} according to \emph{Definition 2}, which is summarized in the below.

  \begin{lemma}
  Given a LM $\mathbf{\eta}$, the optimization problem in \emph{Problem 1} is transformed into the unconstrained optimization problem. Then, the equivalent Bellman equation $(\forall(\mathbf{Q},\mathbf{E})\in \mathcal{Q} \times \mathcal{E})$ is given by
  \begin{align}
  \label{equ:v2}
  \theta+V(\mathbf{Q},\mathbf{E})\nonumber\\
  = \max\limits_{\Omega(\mathbf{Q},\mathbf{E})} \{&g(\eta, (\mathbf{Q},\mathbf{E}), \Omega(\mathbf{Q}, \mathbf{E})) \nonumber \\
  + \sum\limits_{(\mathbf{Q}',\mathbf{E}')}&\Pr[(\mathbf{Q}',\mathbf{E}')|(\mathbf{Q},\mathbf{E}), \Omega(\mathbf{Q},\mathbf{E})] \nonumber \\
  &\times V(\mathbf{Q}',\mathbf{E}') \},
  \end{align}
  where $V(\mathbf{Q},\mathbf{E})= \mathbb{E}[V(\mathbf{H}, \mathbf{Q},\mathbf{E})|(\mathbf{Q},\mathbf{E})] = \sum_{\mathbf{H}\in \mathcal{H}} \Pr[\mathbf{H}]V(\mathbf{H}, \mathbf{Q},\mathbf{E}) $ is the conditional expectation of value function $V(\mathbf{S})$. 
  $\Pr[(\mathbf{Q}',\mathbf{E}')|(\mathbf{Q},\mathbf{E}), \Omega(\mathbf{Q},\mathbf{E})] \!= \mathbb{E}[\Pr[(\mathbf{Q}', \mathbf{E}')| \mathbf{S}, \Omega(\mathbf{S})]|(\mathbf{Q},\mathbf{E})]$ is the conditional expectation of state transition probability, and $g(\eta, (\mathbf{Q},\mathbf{E}), \Omega(\mathbf{Q}, \mathbf{E}))\! = \mathbb{E}[g(\eta, \mathbf{S}, \Omega(\mathbf{S}))|(\mathbf{Q},\mathbf{E})]$ is the conditional per-stage reward.
  \hfill $\blacksquare$
  \end{lemma}

  \begin{remark}
  The equivalent Bellman equation in (\ref{equ:v2}) focus on the $(\mathbf{Q}, \mathbf{E})$ only. However, by solving (\ref{equ:v2}), a stationary subchannel allocation policy $\Omega_c^\ast$ can be attained, which is a function of $(\mathbf{H}, \mathbf{Q}, \mathbf{E})$. Then, the power control policy is determined locally when the action for subchannel allocation is given.   \hfill $\blacksquare$
  \end{remark}

  \section{Low-complexity Solution}
  Owing to the high dimensionality of the state space, it is still very complex to obtain the optimal solution in \eqref{equ:v2}. Brute-force solution brings high computational complexity. Hence, it is preferable to get a low-complexity solution.
  In this section, the subchannel allocation \emph{Q-factor} is approximately converted into the sum of per-slice \emph{Q-factor}. Then, a distributed online learning algorithm is proposed to optimize the per-slice \emph{Q-factor} and the LM of per-slice.

  \subsection{Linear Approximation of the Subchannel Allocation Q-Factor}
  To facilitate the subchannel allocation, we consider the subchannel allocation \emph{Q-factor} $\mathbb{Q}(\mathbf{Q}, \mathbf{E}, \mathbf{c})$ w.r.t. the subchannel allocation action $\mathbf{c}$. According to \emph{Lemma 1}, the optimal subchannel allocation is described in the following corollary.

  \begin{corollary}
  The optimal subchannel allocation is derived as ($\forall(\mathbf{Q},\mathbf{E})\!\in \mathcal{Q} \!\times \mathcal{E}$)
  \begin{align}
  \Omega_c^*(\mathbf{Q},\mathbf{E})\! = \arg \max\limits_{c\in \mathcal{C}}\mathbb{Q}(\mathbf{Q},\mathbf{E},\mathbf{c}),
  \end{align}
  where $\mathbb{Q}(\mathbf{Q},\mathbf{E}, \mathbf{c})$ is the subchannel allocation \emph{Q-factor}. The Bellman equation w.r.t. $(\theta, \mathbb{Q}(\mathbf{Q},\mathbf{E}, \mathbf{c}))$ is given by
  \begin{align}
  \label{equ:q}
  \theta+\mathbb{Q}(\mathbf{Q},\mathbf{E}, \mathbf{c})\nonumber\\
  = \max\limits_{\Omega_p(\mathbf{Q},\mathbf{E})} &\{g(\eta, (\mathbf{Q},\mathbf{E}), \mathbf{c}, \Omega_p(\mathbf{Q}, \mathbf{E})) \nonumber \\
  &+ \sum\limits_{(\mathbf{Q}',\mathbf{E}')}\Pr[(\mathbf{Q}',\mathbf{E}')|(\mathbf{Q},\mathbf{E}),\mathbf{c}, \Omega_p(\mathbf{Q},\mathbf{E})] \nonumber \\
  &\times\max\limits_{\mathbf{p}'\in \mathcal{P}}\mathbb{Q}(\mathbf{Q}',\mathbf{E}', \mathbf{c}') \},
  \end{align}
  where $\Pr[(\mathbf{Q}',\mathbf{E}')|(\mathbf{Q},\mathbf{E}),\mathbf{c}, \Omega_p(\mathbf{Q},\mathbf{E})] = \mathbb{E}[\Pr[(\mathbf{Q}', \mathbf{E}')| \mathbf{S},\mathbf{c}, \Omega_p(\mathbf{S})]|(\mathbf{Q},\mathbf{E})]$ is the conditional expectation of state transition probability, and $g(\eta, (\mathbf{Q},\mathbf{E}), \mathbf{c}, \Omega_p(\mathbf{Q}, \mathbf{E})) = \mathbb{E}[g(\eta, \mathbf{S}, \mathbf{c}, \Omega_p(\mathbf{S}))|(\mathbf{Q},\mathbf{E})]$ is the conditional per-stage reward.
  \end{corollary}

  To further reduce the computational complexity and the cardinality of state space, the \emph{Q-factor} in \eqref{equ:q} is approximated by a linear approximation
  \begin{align}
  \label{equ:linear}
  \mathbb{Q}(\mathbf{Q}, \mathbf{E}, \mathbf{c}) \approx \sum\limits_{m=1}^{M} \hat{\mathbb{Q}}_{ m}(\mathbf{Q}_m, \mathbf{E}_m, \mathbf{c}_{m} ), 
  \end{align}
  where $\mathbf{c}_{m}$ denotes the per-slice subchannel allocation actions of the slice $m$, and $ \hat{\mathbb{Q}}_{ m}(\mathbf{Q}_m, \mathbf{E}_m, \mathbf{c}_{m} )$ is the per-slice \emph{Q-factor} for the slice $m$ of the slice QSI (SQSI) and slice ESI (SESI) $(\mathbf{Q}_m, \mathbf{E}_m)$ and action $\mathbf{c}_{m}$. In addition, the per-slice \emph{Q-factor} is the solution of per-slice fixed-point equation, which is given by
  \begin{align}
  \label{equ:q2}
  \theta_{m} &+ \hat{\mathbb{Q}}_{m}(\mathbf{Q}_m, \mathbf{E}_m, \mathbf{c}_{m})\nonumber\\
  &= \hat{g}_{m}(\mathbf{\eta}_m, \mathbf{Q}_m, \mathbf{E}_m, \mathbf{c}_{m}) \nonumber \\
  &\;\;\;\;+\sum\limits_{(\mathbf{Q}'_m, \mathbf{E}'_m)}\!\Pr\left[(\mathbf{Q}'_m, \mathbf{E}'_m)|(\mathbf{Q}_m, \mathbf{E}_m), \mathbf{c}_{m}\right] \nonumber \\
  &\;\;\;\;\;\;\;\;\;\;\;\;\;\;\;\times \mathbb{E}\left [ \max\limits_{\Omega_{\mathbf{p}_m}} [\hat{\mathbb{Q}}_{m} (\mathbf{Q}'_m, \mathbf{E}'_m, \mathbf{c}'_{m})]\right],
  \end{align}
  where
  \begin{align}
  &\bar{g}_{m}(\eta_m, \mathbf{Q}_{m}, \mathbf{E}_{m}, \mathbf{c}_{m}) \nonumber \\
  &= \mathbb{E}\left[\left( \sum\limits_{i = 1}^{{k_{m}}} \sum\limits_{n = 1}^{N}\omega_m {c^n_{i,m}}B{\log _2}\left( {1 + {\frac{{P_{i,m}|{h_{i,m}^n}{|^2}}} {{\sigma_n^2 }}}} \right) \right. \right. \nonumber \\
  &\left. \left. \;\;\;\;\;\;\;\;+ \eta_m \! \left(\sum\limits_{i = 1}^{{k_{m}}} \overline{D_{i,m}}\!-\!D_m^{\max}\right)\right) | \mathbf{Q}_{m}, \mathbf{E}_{m}, \mathbf{c}_{m}\right],
  \end{align}
  \begin{align}
  &\!\!\Pr[(\mathbf{Q}'_{m},\mathbf{E}'_{m})|(\mathbf{Q}_{m}, \mathbf{E}_{m}), \mathbf{c}_{m}] \nonumber \\
  &\!\!\!= \mathbb{E}[\Pr[\mathbf{Q}'_{m},\mathbf{E}'_{m}|\mathbf{H}_{m}, \mathbf{Q}_{m}, \mathbf{E}_{m}, \mathbf{c}_{m}]|\mathbf{Q}_{m},\mathbf{E}_{m},\mathbf{c}_{m}].
  \end{align}

  Based on the linear approximation of the subchannel allocation \emph{Q-factor} by the sum of per-slice subchannel allocation \emph{Q-factor} in (\ref{equ:linear}), the resource manager decides the optimal subchannel allocation according to
  \begin{align}
  \label{Omega_c}
  \Omega_c^\ast(\mathbf{Q}, \mathbf{E}) = \arg\max\limits_c \sum\limits_{m=1}^{M} \hat{\mathbb{Q}}_{ m}(\mathbf{Q}_m, \mathbf{E}_m, \mathbf{c}_{m} ).
  \end{align}

  Due to the power control can be obtained locally based on the subchannel allocation action.
  According to the slice CSI (SCSI), SQSI and SESI $(\mathbf{H}_m, \mathbf{Q}_m, \mathbf{E}_m)$, the resource manager determines $\Omega_p^\ast(\mathbf{Q}_m, \mathbf{E}_m) = \{\Omega_p^\ast(\mathbf{H}_m, \mathbf{Q}_m, \mathbf{E}_m): (\mathbf{H}_m, \mathbf{Q}_m, \mathbf{E}_m)\forall \mathbf{H}_m\}$, which gets the maximum of the R.H.S. of \eqref{equ:q2}. Therefore, the whole power control policy is attained by $\Omega_p^\ast(\mathbf{S}) = \{\Omega_{\mathbf{p}_m}^\ast((\mathbf{H}_m, \mathbf{Q}_m, \mathbf{E}_m): \mathbf{c}_n \in \Omega_c^\ast(\mathbf{Q}, \mathbf{E})\}$.

  \subsection{Online Learning Algorithm via Stochastic Approximation}
  To solve the (\ref{equ:q2}) and get the LM for each slice, the per-slice delay constraints need to be satisfied, since the control policy $\Omega^\ast = (\Omega_c^\ast, \Omega_p^\ast)$ is function of per-slice \emph{Q-factor} $\{\hat{\mathbb{Q}}(\mathbf{Q}_m, \mathbf{E}_m, {\mathbf{c}_m})\}$.
  In this section, the per-slice \emph{Q-factors} are distributedly estimated by stochastic learning \cite{sl} based on the SCSI, SQSI and SESI, and the LMs also be estimated for each slice.
  The Distributed Online Learning Algorithm using Stochastic Approximation needs to obtain according to the SCSI, SQSI and SESI, which is summarized in the following.

  \emph{Algorithm 1 (Distributed Online Learning Algorithm using Stochastic Approximation):}
  \begin{itemize}
    \item \textbf{Step 1 [Initialization]}: Let $t = 0$. Each slice manager initialize per-slice \emph{Q-factor} $\{\mathbb{Q}_m^0(\mathbf{Q}_m, \mathbf{E}_m, \mathbf{c}_m)\}$, and the LM $\eta_m^0$.
    \item \textbf{Step 2 [Subchannel Allocation]}: Based on SCSI, SQSI and SESI and LM, each slice manager obtain the channel allocation $\mathbf{c}(t) = \{c_{i,m}^n(t)\}_{i\in k_m, m\in M, n\in N}\in \mathcal{C}$ at the beginning of the $t$th slot.
    \item \textbf{Step 3 [Power Control]}: Based on SCSI, SQSI, SESI and LM, each slice manager implements the power control $\mathbf{p}_m(t)= \{\phi_{i,m}(t)\}_{i\in k_m, m\in M} \in \mathcal{P}$ on the basis of $\Omega_{p_m}^\ast(\mathbf{H}_m(t), \mathbf{Q}_m(t), \mathbf{E}_m(t))$.
    \item \textbf{Step 4 [Update Per-Slice Q-Factors and LMs]}:
        Each slice updates per-slice \emph{Q-factor} on the basis of the real-time SCSI, SQSI and SESI after the control action is determined according to \eqref{up_q}. Furthermore, the LMs $\eta_m(t)$ for each slice \emph{m} is updated
        to $\eta_m(t+1)$ according to the iterative formula given by \eqref{up_eta}.
    \item \textbf{Step 5 [Termination]}:  If $||\hat{\mathbb{Q}}(t+1)- \hat{\mathbb{Q}}(t+1) ||< \mu_Q$ and $||\mathbf{\eta}(t+1)- \mathbf{\eta}(t)|| < \mu_{\eta}$, stop; otherwise, set $t := t + 1$ and go to step 2.
  \end{itemize}

  The per-slice \emph{Q-factor} update in Step 4 according to SCSI, SQSI and SESI at the current time slot is given by
  \begin{align}
  \label{up_q}
  &\hat{\mathbb{Q}}^{t+1}_m(\mathbf{Q}_m, \mathbf{E}_m, \mathbf{c}_m) \nonumber \\
  &= \hat{\mathbb{Q}}^t_m(\mathbf{Q}_m, \mathbf{E}_m, \mathbf{c}_m) + \epsilon_t [U_m(\hat{\mathbb{Q}}^{t}_m, \mathbf{Q}_m, \mathbf{E}_m, \mathbf{c}_m) \nonumber \\
  &- U_m(\hat{\mathbb{Q}}^{t}_m, \mathbf{Q}_m^I, \mathbf{E}_m^I, \mathbf{c}_m^I)-\hat{\mathbb{Q}}^{t}_m(\mathbf{Q}_m, \mathbf{E}_m, \mathbf{c}_m)],
  \end{align}
  where
  \begin{align}
  &U_m(\hat{\mathbb{Q}}^{t}_m, \mathbf{Q}_m, \mathbf{E}_m, \mathbf{c}_m) \nonumber \\
  &= \hat{g}_m(\mathbf{Q}_m, \mathbf{E}_m, \mathbf{c}_m) \nonumber \\
  &+ \sum\limits_{(\mathbf{Q}'_m,\mathbf{E}'_m)} \Pr[(\mathbf{Q}'_m, \mathbf{E}'_m)| (\mathbf{Q}_m, \mathbf{E}_m), \mathbf{c}_m] \nonumber \\
  &\times \mathbb{E}\left[ \mathbb{Q}_m^t\left( \min\{Q_m^t + A_m^Q(t), B_m^{\rm{Q}}\}, \right. \right. \nonumber \\
  &\left. \left. \min\{E_m^t + A_m^E(t), B_m^{\rm{E}}\}, \mathbf{c}'_{m}\right)\right].
  \end{align}

  The LMs $\eta_m(t)$ for each slice $m$ can be updated as
  \begin{align}
  \label{up_eta}
  \eta_m(t\!+1) \!= \Gamma(\eta_m(t\!+1)\!+\epsilon_{\eta}(t)(\bar{D}_m(t)\!-D_m^{\max})).
  \end{align}

  \subsection{Decomposition of the Per-slice Q-factor}
  Note that the cardinality is $(N_Q+1)^{k_m}(N_E+1)^{k_m}$ for the per-slice system state, which is exponential in the number of UEs at slice $m$th , i.e., $k_m$. We point out that the per-slice \emph{Q-factor} can be further split into per-slice per-UE \emph{Q-factor} in the following lemma, which results in a linear order for the state space cardinality, i.e., $k_m(N_Q+1){k_m}(N_E+1)$.

  \begin{lemma}
  \emph{(Decomposition of Per-slice Q-factor):}
  The per-slice \emph{Q-factor} $\hat{\mathbb{Q}}_{m}(\mathbf{Q}_m, \mathbf{E}_m, \mathbf{c}_{m})$ can be decomposed into the sum of the per-slice per-UE \emph{Q-factors} $\hat{\mathbb{Q}}_{i,m}(Q_{i,m}, E_{i,m}, \mathbf{c}_{i,m})$, i.e., $ \hat{\mathbb{Q}}_{m}(\mathbf{Q}_m, \mathbf{E}_m, \mathbf{c}_{m}) = \sum\limits_{i = 1}^{{k_{m}}} \hat{\mathbb{Q}}_{i,m}(Q_{i,m}, E_{i,m}, \mathbf{c}_{i,m})$, where $\hat{\mathbb{Q}}_{i,m}(Q_{i,m}, E_{i,m}, \mathbf{c}_{i,m})$ satisfies the per-slice per-UE \emph{Q-factor} equation in \eqref{perue}.
  \begin{figure*}[ht]
  \begin{align}
  \label{perue}
  &\theta_{i,m} + \hat{\mathbb{Q}}_{i,m}(Q_{i,m},E_{i,m},\mathbf{c}_{i,m})\nonumber\\
  &\;\;\;= \max\limits_{\Omega_{p_{i,m}}} \{\hat{g}_{i,m}(\mathbf{\eta}_m, (Q_{i,m},E_{i,m}), \mathbf{c}_{i,m}) +\sum\limits_{(Q'_{i,m},E'_{i,m})}\Pr[(Q'_{i,m},E'_{i,m})|(Q_{i,m},E_{i,m}), \mathbf{c}_{i,m}] W_{i,m}(Q'_{i,m},E'_{i,m}) \},
  \end{align}
  \hrulefill
  \end{figure*}
  In \eqref{perue}, $\hat{g}_{i,m}(\eta_{i,m}, Q_{i,m},E_{i,m}, \mathbf{c}_{i,m})$ and $\Pr[(Q'_{i,m},E'_{i,m}|(Q_{i,m},E_{i,m}), \mathbf{c}_{i,m}]$ are given by
  \begin{align}
  \label{sumn}
  &\hat{g}_{i,m}(\eta_{i,m}, Q_{i,m},E_{i,m}, \mathbf{c}_{i,m}) \nonumber \\
  &= \mathbb{E}\left[ \left( \sum\limits_{n = 1}^{N} \omega_m{c^n_{i,m}}B{\log_2}\left( {1 + {\frac{{P_{i,m}|{h_{i,m}^n}{|^2}}}{ {\sigma_n^2 }}}} \right) \right. \right. \nonumber \\
  &\left. \left. + \eta_m (D_{i,m}-{\frac{D_m^{\max}}{ {k_m}}})\right) |Q_{i,m},E_{i,m}\right],
  \end{align}
  and
  \begin{align}
\Pr&[(Q'_{i,m},E'_{i,m}|(Q_{i,m},E_{i,m}),\mathbf{c}_{i,m})] \nonumber \\
  &= \mathbb{E}[\Pr[Q'_{i,m},E'_{i,m}|\mathbf{H}_{m}, Q_{i,m},E_{i,m}, \nonumber\\
  &\Omega_{c_{i,m}}(\mathbf{H}_{m}, Q_{i,m},E_{i,m})]| Q_{i,m},E_{i,m}].
  \end{align}

  \hfill $\blacksquare$
  \end{lemma}

  Recall that the subchannel $n\in \{1, 2, \ldots, N\}$ can be allocated to at most one UE. In addition, observe that the summation index \emph{n} in \eqref{sumn} is the index of subchannels. Consequently, for every \emph{n}, at most one $c^n = 1$, while all the other $c^n = 0$.
  Due to the birth-death queue dynamics and symmetry of each subchannel \cite{dab}, the per-slice per-UE \emph{Q-factor} meeting \eqref{perue} can be converted into the sum of the per-slice per-UE per-subchannel \emph{Q-factors}, which can be written as
  \begin{align}
  \!\!\!\hat{\mathbb{Q}}_{i,m}\!(Q_{i,m}, E_{i,m},\mathbf{c}_{i,m}\!) \!= \!\sum\limits_{n = 1}^{N}\! \hat{\mathbb{Q}}_{i,m}\!(Q_{i,m}, E_{i,m}, c_{i,m}^n),
  \end{align}
  where per-slice per-UE per-subchannel \emph{Q-factor} satisfies:
  \begin{align}
  &\bar{g}_{i,m}(\eta_{i,m}, Q_{i,m},E_{i,m}, \Omega_{c^n_{i,m}}(Q_{i,m},E_{i,m})) \nonumber \\
  &= \mathbb{E}\left[ \left( \omega_m{c^n_{i,m}}B{\log_2}\left( {1 + {\frac{{P_{i,m}|{h_{i,m}^n}{|^2}}}{ {\sigma_n^2 }}}} \right) \right. \right. \nonumber \\
  &\left. \left. + {\frac{\eta_m }{ N}} (D_{i,m}-{\frac{D_m^{\max}}{{k_m}}})\right) |Q_{i,m},E_{i,m}\right].
  \end{align}

  Then, the subchannel allocation action is determined as
  \begin{equation}
  c_{i,m}^n = \left\{ {\begin{array}{*{20}{l}}
{1,\;{\text{if}}\;{F} > {\max}_j {F_j} }\\
{0,\;{\text{otherwise}}}
\end{array}} \right.,
  \end{equation}
  where $F = \hat{\mathbb{Q}}_{i,m}\!(Q_{i,m}, E_{i,m}, c_{i,m}^n)$ and $F_j = \hat{\mathbb{Q}}_{i,m}\!(Q_{i,m}, E_{i,m}, c_{i,m}^j)$.

  \section{Simulation Results}
  In this section, the proposed optimal and reduced complexity solutions is compared with three reference baselines to demonstrate the advantage of the proposed scheme.
  \begin{itemize}
    \item \textbf{Baseline 1 [Random Control Scheme]}: The virtual resources are assigned to each slice equally. On the basis of this policy, each slice can utilize the radio resources with the same opportunity. In this scheme, the UE always transmit with the maximum power.
    \item \textbf{Baseline 2 [QSI-Based Scheme]}: In order to optimize the weighted-sum transmission rate, the subchannel allocation and power control policy are determined according to the CSI and QSI.
    \item \textbf{Baseline 3 [Value Iteration Scheme]}: The value iteration scheme can find the optimal policy and value functions. 
        Hence, it is regarded as the upper bound of performance.
  \end{itemize}

  \begin{table}[b]
  \centering
  \caption{The parameters for RAN}
  \begin{tabular}{|c|c|}
    \hline
    Parameters       & Values                                    \\  \hline
    Radius      & 100 m                                     \\  \hline
    Background noise power ($\sigma^2$) & -104 dBm     \\  \hline
    Subchannel Bandwidth & 8 MHz     \\  \hline
    Number of subchannels & 6                        \\  \hline
    FPC factor             & 0.6, 0.8, 1                                               \\  \hline
  \end{tabular}
  \end{table}

   \begin{table}[b]
  \newcommand{\tabincell}[2]{\begin{tabular}{@{}#1@{}}#2\end{tabular}}
  \centering
  \caption{The parameters for three kinds of slices}
  \begin{tabular}{|c|c|c|c|}
    \hline
    Parameters       & eMBB  & URLLC  & mMTC                                   \\  \hline
    Number of UEs   & 2 & 2  & 2                        \\  \hline
    Baseline power   & -76 dBm  & -73 dBm & -79 dBm                        \\  \hline
    Queue state  & 6    & 6     & 4 \\  \hline
    Energy state & 6    & 6    & 4  \\  \hline
    Packet size  & 100 kbits        & 10 kbits     & 2 kbits                                \\  \hline
    \tabincell{c}{Average packet \\ arrival rate}   & 3 packet/$\Delta$   & 3 packet/$\Delta$   & 3 packet/$\Delta$         \\  \hline
    \tabincell{c}{Average energy \\ arrival rate}   & 3 unit/$\Delta$     & 3 unit/$\Delta$   & 3 unit/$\Delta$                 \\  \hline
    Maximum delay  & 100 ms        & 10 ms     & 3 s                                \\  \hline
  \end{tabular}
  \end{table}

  Consider that a RAN coverage area with radius 100m has three types of network slices to support URLLC, eMBB and mMTC, respectively. And the mean of fading is fixd to $\bar{g} = -10$ dB. The main parameter values of the RAN are listed in Table II.
  Each slice supports a kind of UE. The main parameter values of the slice UEs are listed in Table III. There are 6 subchannels with bandwidth 8MHz. A data queue is maintained at each UE to cache the arrival packets.
  The UE keeps a queue to cache the arrival data packets. The UEs in different slices have different buffer size and battery capacity.

  Fig. 3 to Fig. 5 compare the performance of the proposed optimization scheme with Random Control Scheme and QSI-Based Scheme.
  Simulation results indicate that the proposed optimization scheme achieves a performance improvement and outperforms the baseline 1 and baseline 2.
  The virtual resources are assigned to each slice equally and the maximum transmission power is adopted in Baseline 1, hence it realizes worse performance. Baseline 2 cares the CSI and QSI and does not take into consideration the ESI, which is better than baseline 1 and worse than the proposed scheme.
  Besides to replacing the equality allocation with on-demand subchannel allocating among UEs, the performance improvement is also owe to the power control policy in the proposed optimization scheme.
  Fig. 6 indicates that the performance of the proposed optimization scheme is asymptotically optimal and highly close to the upper bound.

  \begin{figure}[t]
   \centering
    \includegraphics[width=2.8in,angle=0]{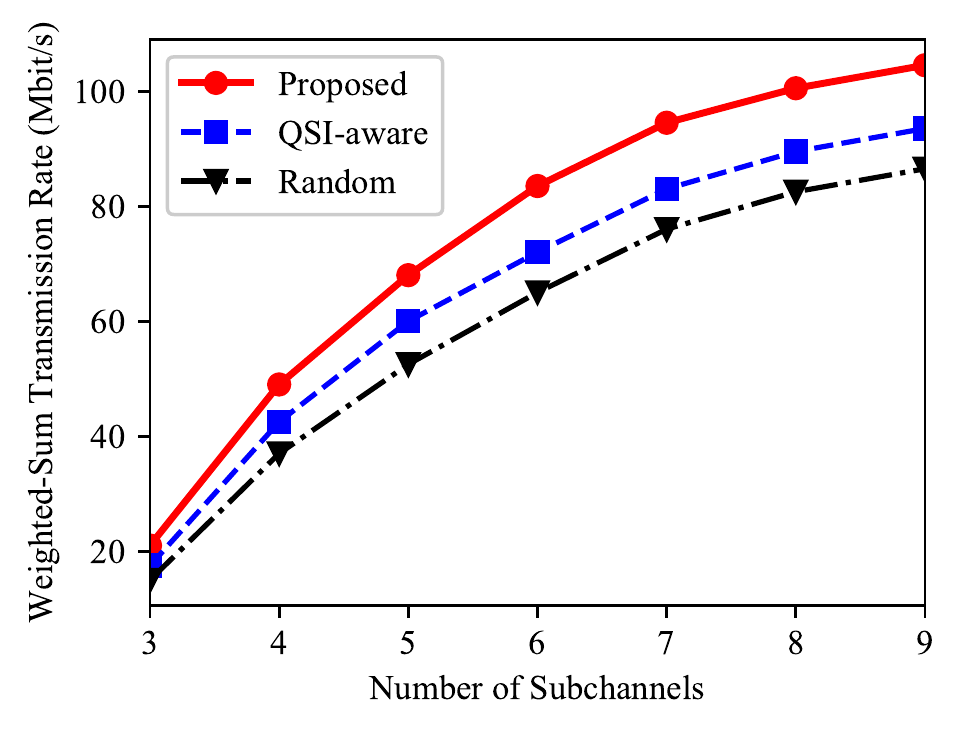}
    \caption{Weighted-sum transmission rate versus the number of subchannels.}
   \end{figure}

   \begin{figure}[t]
   \centering
    \includegraphics[width=2.8in,angle=0]{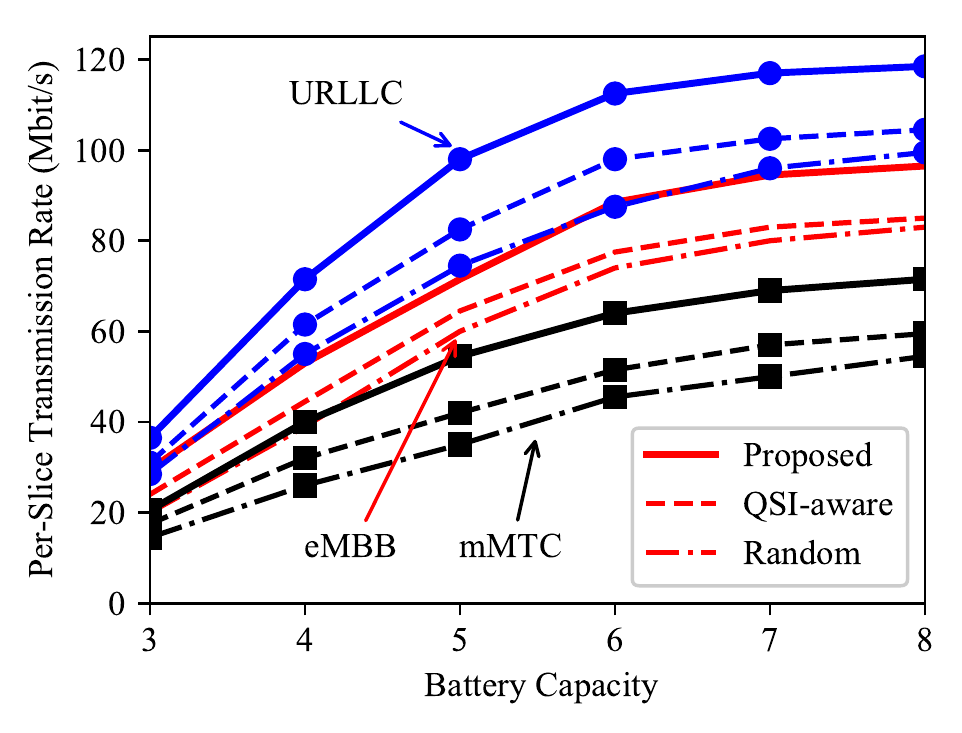}
    \caption{Per-slice transmission rate versus battery capacity.}
\end{figure}

  \begin{figure}[t]
   \centering
    \includegraphics[width=2.8in,angle=0]{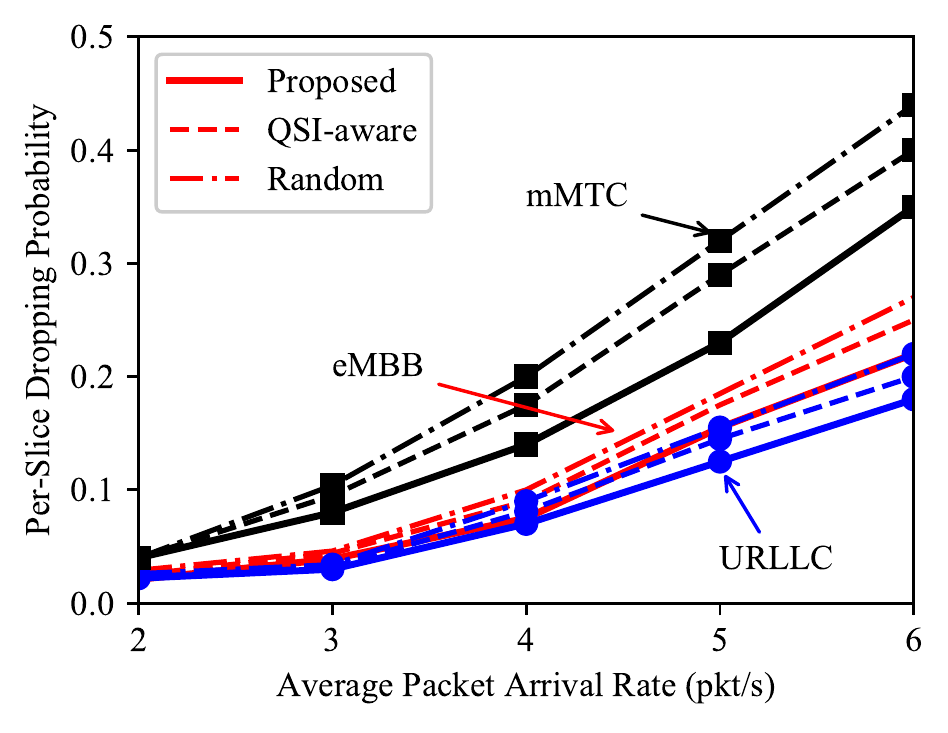}
    \caption{Per-slice dropping probability versus the average packet arrival rate.}
   \end{figure}

  Fig. 3 illustrates weighted-sum transmission rate versus the number of subchannels.
  For all the schemes, we can observed that weighted-sum transmission rate increases with the number of subchannels.
  Due to the lack of channel supply at the beginning, only some UEs get the opportunities to transmit packets, which results in the number of packets being dropped and the number of packets remaining in the queue.
  Even if every user can get the channel access opportunities, in order to reduce both the packet drops and delay, the slice manager may need to schedule more queued packets for transmission when the channel environment is poor, therefore the higher transmission power will be selected.
  As the number of channels increases, multiple subchannels may be allocated to one UE. At this time, the transmission rate of UEs grow slowly as a result of the limitation of buffer size and average packet arrival rate.

  Fig. 4 illustrates per-slice transmission rate versus battery capacity. For all the schemes, The simulation results show that the per-slice transmission rate increases with the battery capacity. As the battery capacity gradually increases, the rate increases slowly. This is because even though the battery capacity increases, the average energy arrival does not increase.
  Moreover, the proposed optimization scheme have significant performance improvement compared with baseline 1 and baseline 2. The transmission rate of URLLC slice with strict delay constraints is higher than that of the other two slices. In addition, mMTC slice has the lowest delay requirement and the smallest packet size, resulting in the lowest transmission rate.
  Observe that the performance gaps between the baseline 1 and baseline 2 become smaller with the increase of battery capacity. This is because more energy is stored in the battery as the battery capacity increases, the slice manager is more inclined to choose the maximum transmit power.

  Fig. 5 illustrates per-slice dropping probability versus the average packet arrival rate.
  As the average packet arrival rate increases, the dropping probability increases gradually.
  Observe that mMTC slice has the largest dropping probability due to the limitation of buffer capacity. The buffer capacity of mMTC slice is 3, and the buffer capacity of eMBB slice and URLLC slice are 6. After the average packet arrival rate is greater than 3, the dropping probability of mMTC slice increases rapidly, the dropping probability of eMBB slice and URLLC slice increase relatively slowly.

  Fig. 6 illustrates the convergence property of the proposed optimization scheme for estimating per-slice \emph{Q-factor} and the value iteration algorithm given in baseline 3.
  It turns out that the proposed optimization scheme converges quite fast.
  The performance at 80-th number of iteration is already quite close to the converged value.

  \begin{figure}[t]
   \centering
    \includegraphics[width=2.8in,angle=0]{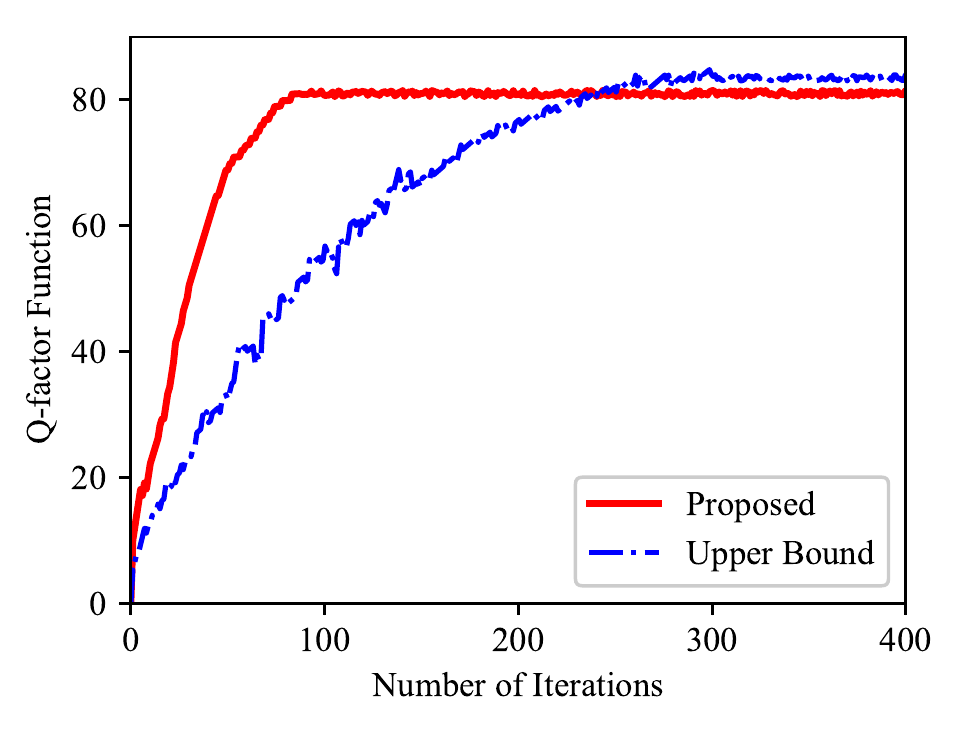}
    \caption{Convergence property of the proposed online stochastic learning algorithm.}
   \end{figure}

  \section{Conclusion}
  In this paper, we propose a dynamic virtual resources allocation scheme based on RAN slicing for uplink communication to guarantee the QoS. The resources control problem is formulated as an infinite-horizon average-reward CMDP problem.
  To reduce the computational complexity, an equivalent Bellman equation on the basis of subchannel allocation \emph{Q-factor} is proposed to solve the CMDP. In addition, we adopt a distributed online stochastic learning algorithm to optimize the value functions and LMs. Then, the subchannel allocation \emph{Q-factor} is approximate to the sum of per-slice \emph{Q-factors} to further reduce the computational complexity. Finally, simulation results reveal that the proposed optimization scheme can converge and improve the user performance compared with baseline 1 and baseline 2.

\bibliographystyle{IEEEtran}
\bibliography{reference}
\end{document}